\documentclass{article}

\usepackage{microtype}
\usepackage{graphicx}
\usepackage{subfigure}
\usepackage{booktabs} 
\usepackage{amsfonts,amssymb}
\usepackage{hyperref}
\usepackage{multirow}
\usepackage{color}
\usepackage{listings}
\usepackage[ruled,linesnumbered]{algorithm2e}

\definecolor{dkgreen}{rgb}{0,0.6,0}
\definecolor{gray}{rgb}{0.5,0.5,0.5}
\definecolor{mauve}{rgb}{0.58,0,0.82}

\newcommand{\tabincell}[2]{\begin{tabular}{@{}#1@{}}#2\end{tabular}}  
\usepackage[accepted]{icml2021}

\icmltitlerunning{How could Neural Networks understand Programs?}

\begin{document}

\twocolumn[
\icmltitle{How could Neural Networks understand Programs?}



\icmlsetsymbol{equal}{*}

\begin{icmlauthorlist}
\icmlauthor{Dinglan Peng}{ustc}
\icmlauthor{Shuxin Zheng}{msra}
\icmlauthor{Yatao Li}{msra}
\icmlauthor{Guolin Ke}{msra}
\icmlauthor{Di He}{msra}
\icmlauthor{Tie-Yan Liu}{msra}
\end{icmlauthorlist}

\icmlaffiliation{ustc}{University of Science and Technology of China}
\icmlaffiliation{msra}{Microsoft Research Asia}

\icmlcorrespondingauthor{Shuxin Zheng, Yatao Li, Di He}{\{shuz,yatli,dihe\}@microsoft.com}

\icmlkeywords{Code Representation, Operational Semantics}

\vskip 0.3in
]



\printAffiliationsAndNotice{}  

\begin{abstract}

Semantic understanding of programs is a fundamental problem for programming language processing (PLP). Recent works that learn representations of code based on pre-training techniques in NLP have pushed the frontiers in this direction. However, the semantics of PL and NL have essential differences. These being ignored, we believe it is difficult to build a model to better understand programs, by either directly applying off-the-shelf NLP pre-training techniques to the source code, or adding features to the model by the heuristic. In fact, the semantics of a program can be rigorously defined by formal semantics in PL theory. For example, the operational semantics, describes the meaning of a valid program as updating the environment (i.e., the memory address-value function) through fundamental operations, such as memory I/O and conditional branching. Inspired by this, we propose a novel program semantics learning paradigm, that the model should learn from information composed of (1) the representations which align well with the fundamental operations in operational semantics, and (2) the information of environment transition, which is indispensable for program understanding. To validate our proposal, we present a hierarchical Transformer-based pre-training model called OSCAR to better facilitate the understanding of programs. OSCAR learns from intermediate representation (IR) and an encoded representation derived from static analysis, which are used for representing the fundamental operations and approximating the environment transitions respectively. OSCAR empirically shows the outstanding capability of program semantics understanding on many practical software engineering tasks. Code and models are released at: \url{https://github.com/pdlan/OSCAR}.

\end{abstract}

\section{Introduction}
Modern software typically contains tons of code, functions, and modules with overwhelmingly complex structure or organization scheme. It poses great challenges for writing, maintaining, and analyzing such programs. Fortunately, a series of deep learning-based productivity tools were developed to automatically help programmers by analyzing program~\cite{ding2019asm2vec,duandeepbindiff,yu2020order}, security auditing~\cite{zhou2019devign,buratti2020exploring}, code retrieval~\cite{luan2019aroma,ye2020misim,cummins2020deep}, and so on. 

Inspired by the success of pre-trained representation for semantics understanding of natural language~\citep{devlin2019bert,brown2020language,xiong2020layer}, there are many recent attempts to graft the conventional NLP pre-training techniques to source code~\citep{buratti2020exploring,feng2020codebert,lachaux2020unsupervised,guo2020graphcodebert,yu2020order}, in which the code representation is obtained by capturing contextual information from a substantial amount of source code text, and is then used for a variety of downstream software engineering tasks after fine-tuning. For instance, CuBERT~\citep{kanade2020learning} leverages the powerful pre-training contextual embedding model BERT~\citep{devlin2019bert} to learn informative representations on a Python corpus; CodeBERT~\cite{feng2020codebert} learns general-purpose representations to bridge natural language (NL) and high-level programming language (PL) by pre-training on NL-PL pairs. Furthermore, features designed by experts (e.g., data flow graph)~\cite{guo2020graphcodebert} are added to the pre-training model, aiming to provide additional information for program semantics understanding.

However, programming languages have many fundamental differences in essence with natural languages. For example, the same program may exhibit different behaviors against its input and memory state, while there is no such explicit concept in natural language. Therefore, we argue that the current approaches that attempt to capture the semantic proprieties directly from the source code, will limit the semantics understanding of programs, be it applying off-the-shelf NLP pre-training techniques, or adding features to the model by the heuristic.

Indeed, the rigorous mathematical account of the meaning (i.e., the semantics) of programming languages~\cite{gunter1992semantics}, has been well-studied by formal semantics~\cite{winskel1993formal} in programming language theory. For instance, the operational semantics~\cite{van2012revised}, which is a widely used branch of formal semantics, captures the meaning of a programming language by defining rules for how its programs execute on an abstract machine. These rules reflect the environment transitions according to the instructions, where the \textit{environment}~\cite{stuart2013understanding} is formally defined as a function mapping all memory addresses to their values\footnote{For simplicity, we consider that all the values are stored in memory, e.g., LLVM values.}, and one \textit{instruction} conducts a rigorously defined operation, such as reading/writing the memory, basic arithmetic, boolean logic, or conditional branching.

Inspired by the programming language theory, we propose a code representation learning paradigm which could make a model better understand programs. In particular, a code representation should be learned from (1) a translation of the source code text that aligns well with those fundamental operations defined in operational semantics; (2) the information of environment transition, which is obviously indispensable for program understanding. 

In order to verify the effectiveness of our proposal, we further present a novel pre-training model called Operational Semantics for Code Abstract Representation (OSCAR) based on a hierarchical Transformer~\cite{vaswani2017attention}, which is designed to capture the contextual information among long sequences of code. On one hand, to represent the fundamental operations, OSCAR utilizes intermediate representation (IR), which is more applicable for learning code representation rather than high-level programming languages, since the IR is modeled on an abstract machine with a finite instruction set, which can be mapped to the operational semantics almost perfectly. In particular, the IR can be easily acquired by translation from binary or source code of a target program. On the other hand, obtaining concrete and precise information of the environment transition requires plenty of actual executions and calculations, which would be impractical and risky. Therefore, OSCAR alternatively uses abstract information, which can be obtained by abstract interpretation inspired static program analysis without difficulty. Abstract interpretation~\cite{cousot1977abstract,cousot1979systematic} describes program semantics by a mathematical characterization of possible behaviors of the program instead of modeling the behaviors after many actual execution trails of the program. In addition, to capture the control structure of a target program or code snippet, we develop a novel Positional Condition Encoding (PCE) to encode the control flow information into the model. 

Furthermore, to ensure the desired capacity of pre-trained representation, we design a compact and effective objective function by combining two respective components: a variant of MLM loss~\cite{devlin2019bert} masking entire instructions, and a contrastive loss with different compilation optimization techniques. With instruction tokens as input, the model could capture token-level contextual knowledge by optimizing the variant of MLM loss. Meanwhile, by generating syntactically diverse but functionally equivalent IRs through different compilation optimization techniques, e.g., strength reduction, loop unrolling, and inline expansion, the contrastive loss could provide informative self-supervision, to help the model to efficiently capture the program- or code snippet-level semantic knowledge.

Our contributions are concluded as follows:
\begin{itemize}
    \item We propose a new learning paradigm that suggests the pre-training model could learn code representation from both the superficial instructions and the underlying environment transitions, which alleviates the aforementioned limitations of semantics understanding of program according to operational semantics.
    \item We demonstrate our proposal by presenting OSCAR, a hierarchical Transformer which represents the fundamental operations by IR and approximates the environment transitions by an encoded representation derived from static analysis. We also design efficient training objectives for OSCAR to largely facilitate the program semantics understanding.
    \item OSCAR significantly boosts the performance of semantics understanding of program on a wide range of downstream practical software engineering tasks. Moreover, OSCAR shows remarkable zero-shot ability, i.e., without fine-tuning the parameters, comparing to state-of-the-art pre-training methods.
\end{itemize}

\section{Related Work}

Inspired by the great success of deep learning on natural language understanding, there is a growing body of exploratory work on programming language understanding by incorporating code structure into DNN, such as abstract syntax tree (AST)~\cite{alon2020structural, rabinovich2017abstract, yin2017syntactic, wei2017supervised, chen2018tree,alon2018code2seq,poj104,alon2019code2vec,zhang2019novel,bui2020infercode} or graph~\cite{brockschmidt2018generative, wang2020detecting, allamanis2018learning, hellendoorn2019global,duandeepbindiff,cummins2020programl,ye2020misim,hellendoorn2019global, david2020neural, wang2020blended}. 

As the most commonly used architectures in NLP, the Transformer~\cite{vaswani2017attention} has also been widely adopted in code understanding tasks.~\citet{kim2020code} achieve high accuracy of next token prediction on code by feeding AST to Transformer.~\citet{svyatkovskiy2020intellicode} propose to train a variant of GPT-2~\cite{radford2019language} from scratch on source code to improve the performance of code completion. Recent works employ pre-training on large-scale code corpus and achieve promising results of code representation.~\citet{kanade2020learning} pre-train a BERT model on a massive corpus of Python source code and get outstanding performance on five code intelligence tasks.~\citet{buratti2020exploring} present C-BERT, which is a transformer-based model pre-trained on a large collection of corpus written in C.~\citet{feng2020codebert} propose a cross-modal BERT called CodeBERT between source codes and comments, written in programming language and natural language respectively, and gain excellent achievements on NL-PL tasks, such as code search by natural language and code generation from comments. By introducing the data flow to the model,~\citet{guo2020graphcodebert} further improve the performance of CodeBERT.  ~\citet{ahmad2021unified} present PLBART, which is also pre-trained on a cross-modal corpus of programming language and natural language via denoising autoencoding. BinaryAI~\cite{yu2020order} leverage a BERT model pre-trained on binary code to construct a hybrid model by combining with GNN and CNN, and achieves excellent performance on binary code similarity detection.~\citet{yu2020codecmr} further introduce a novel CNN as a feature extractor for source code, and improve the performance of binary-source code matching.

To our best knowledge, the proposed OSCAR is the first attempt for code representation using our PL theory-inspired learning strategy that considers both the superficial programming language and the underlying environment transitions, to improve the performance of program and code understanding.

\textbf{Intermediate Representation} \quad There are prior works~\cite{ben2018neural,venkatakeerthy2020ir2vec,cummins2020deep} that attempt to understand code on IR language with different motivations. For example,~\citet{ben2018neural} argues that training model on a specific source programming language (or machine code for optimization) could not generalize to other languages, and suggests that training on IR language is better since it accepts code in various source languages. Similarly,~\citet{cummins2020deep} aims to produce a language-agnostic, compiler-independent representation for the program by leveraging a corpus of LLVM IR covering six source programming languages. 

Different from the motivations of previous methods, we suggest that the IR is more applicable for learning code representation rather than high-level PL since the IR is modeled on an abstract machine with a finite instruction set, which can be well mapped to operational semantics.

\textbf{Contrastive Learning} \quad In recent years, contrastive learning~\cite{hadsell2006dimensionality,chen2020simple,he2020momentum} shows promising results on unsupervised visual representation learning. Inspired by this,~\citet{jain2020contrastive} present ContraCode, which applies contrastive learning on code representation learning by adopting several source-to-source transformations, such as variable renaming, dead code elimination, and dead code insertion. The functionality of the program would not be changed after the transformations, therefore the underlying representations should be the same.

Different from ContraCode, we generate syntactically diverse but functionally equivalent IRs with different optimization techniques in compilers. Unlike the transformations in ContraCode, different optimizations would lead to huge differences in the syntax and structure of the IR, such as strength reduction, loop unrolling, and inline expansion. This kind of diversity could provide informative self-supervision in helping the model to efficiently capture the program- or code snippet-level semantic knowledge.

\section{Method}

\begin{figure*} [!htb]
    \centering
    \includegraphics[scale=0.6]{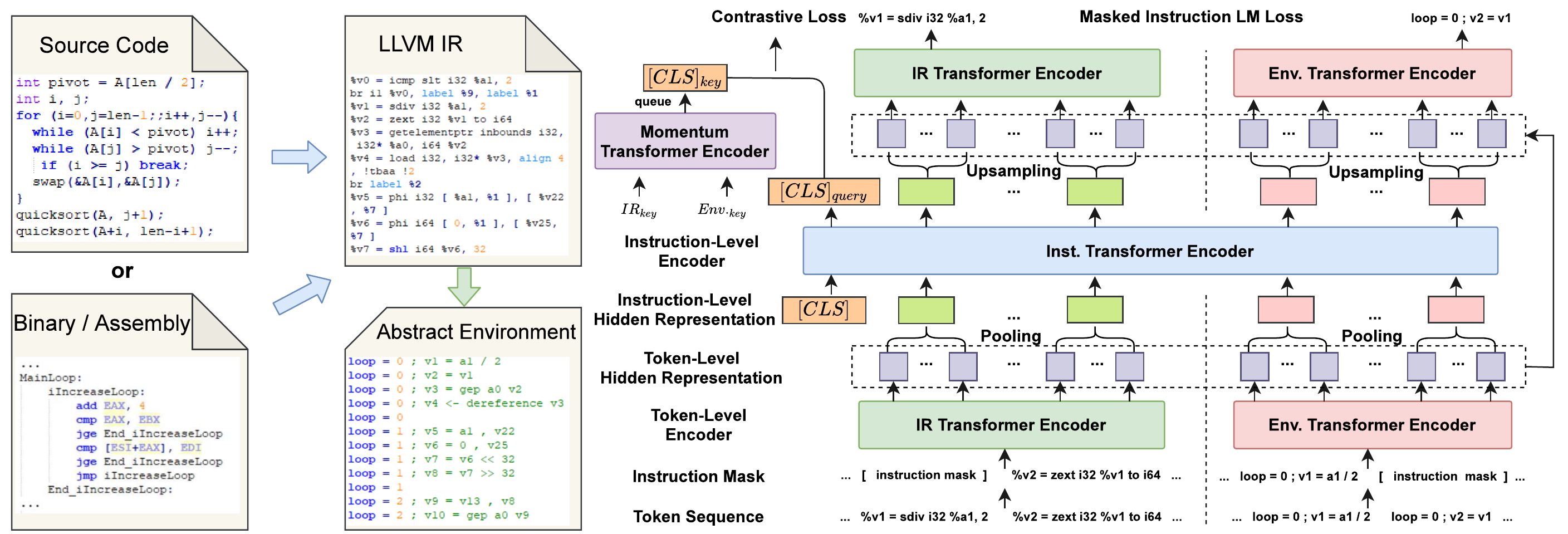}
    \caption{\textbf{An illustration of the model architecture of OSCAR.}}
    \label{fig:cros}
\end{figure*}

As mentioned above, operational semantics~\cite{van2012revised, stuart2013understanding} captures the meaning of an executable program by the environment transitions according to the instructions on an abstract machine. To be more concrete, we illustrate our motivation by structural operational semantics~\cite{plotkin1981structural, hennessy1990semantics}. The meaning of assignment and composition on a simplified abstract machine can be represented respectively as
$$
\frac{\langle E,s \rangle \Longrightarrow V}{ \langle L:=E,s \rangle \rightarrow ( s\uplus  (L \mapsto V) )   }, \quad  \frac{ \langle C_1,s \rangle \longrightarrow s'}{\langle C_1;C_2,s \rangle \rightarrow \langle C_2,s' \rangle},
$$
where $E,L,V$ denote expression, memory location and value respectively, $s\in \mathcal{S}$ denotes environment function mapping all memory locations to values, and $C$ represents code snippet. Therefore, the meaning of assignment can be explained as ``the program $L:=E$ will update the environment function $s$ with $L=V$ if the expression $E$ in the environment $s$ reduces to $V$''. Similarly, the composition can be explained as ``if the code snippet $C_1$ in environment $s$ finishes in $s'$, then the composed code snippet $C_1;C_2$ in environment $s$ can reduce to execute $C_2$ in $s'$''. 

Obviously, the semantics of a code snippet depends on two parts: the instructions and the information of environment transitions on the abstract machine. Therefore, we propose that a good code representation would be sufficiently learned from these two parts to better understand the semantics of programs. In the following, we will present OSCAR, which is a hierarchical model that learns code representation from these two aspects.

\subsection{Input Representations}
\subsubsection{Intermediate Representation (IR)}

Learning representation directly from high-level PLs has been widely adopted by existing program understanding methods. However, the gap between the textual representation of source or binary code versus the actual computational meaning becomes larger along with the development of modern programming languages and compilers. This non-negligible gap increases the difficulty of code understanding for existing models.

In general, in order to better analyze and optimize a program, a modern compiler will translate the source code into IR before it generates machine code for a target architecture. IR is modeled after an abstract machine that is typically designed so that each instruction represents exactly one fundamental operation. With this characteristic, IR becomes a more accurate and appropriate representation of the instruction in operational semantics instead of high-level PLs. We collect a large corpus of real-world programs (Details in Appendix F.1) and translate them into LLVM IR as our pre-training data. LLVM IR is one of the most commonly used IR forms, and supports a wide variety of languages.

There is an additional advantage for using IR: if the target code snippet is binary or assembly, the textual information would be easily preserved when translating binary code to IR, unless the binary is generated from strong obfuscation, e.g., executable packing. Meanwhile, translating binary or assembly back to source code (aka. decompilation) would totally change the distribution and legibility of tokens, which would deeply hurt the performance of source code-based methods.

\subsubsection{Abstract Environment Information}

We leverage the structural operational semantics to illustrate how we encode the information of environment transitions into the model.
The inductive nature of structural operational semantics requires a properly defined initial condition, which is described by the initial environment function. The transitions can then be inferred step-by-step based on the sequencing rules (i.e., composition, transformation, and conditioning, please refer to ~\citet{plotkin1981structural,hennessy1990semantics}). To fully capture the concrete and precise information of environment transitions, one has to iterate through many possible combinations of input values and initial conditions, and infer the transitions by actually execute the program with the sequencing rules. This is obviously infeasible since actual executions are quite time-consuming and risky, e.g., analysis for large software projects or malicious software.

Therefore, we alternatively use the abstract environment information obtained from static program analysis, instead of the concrete one. The abstract environment information is inspired by the abstract interpretation~\cite{cousot1977abstract,cousot1979systematic}, and describes program semantics by a mathematical characterization of possible behaviors of the program instead of modeling the behaviors after many actual execution trails of the program. Applying this idea to structural operational semantics, each expression can reduce to not only a concrete value, but also a relation or a possible range in the value space. 

Specifically, we extract three types of relational constraints of the environment from the instructions: those governed by static single assignment (SSA), those by memory reads, and those by memory writes. This information can be easily obtained by LLVM built-in analytic features, e.g., \textit{MemorySSA}. In addition, to better model the range constraints of the environment
, we extract auxiliary information from the control flow graph, i.e., the depth of loop, via LLVM \textit{LoopInfo}. Detailed descriptions about the extraction of abstract environment information can be found in Appendix A.

\subsection{Model}

\subsubsection{Architecture}

The model architecture of OSCAR is a hierarchical multi-layer Transformer encoder, which is illustrated in Fig.\ref{fig:cros}. In particular, OSCAR consists of two levels of encoders. The lower level is composed of two token-level encoders, which are used to process tokens from IR and abstract environment information, respectively. The upper level is an instruction-level encoder, which aims to extract features further based on the lower-level layer's outputs. The implementation of each level of encoders is identical to BERT~\cite{devlin2019bert}. We call the two token-level encoders as IR and Env. encoder, and the instruction-level encoder as Inst. encoder. 

Typically, the token sequence of a practical program is long. If we simply feed the sequence to a standard Transformer, the time and space cost will be extremely high since the attention module suffers from quadratic computation and memory requirements with respect to the sequence length. Most previous methods truncate the long input sequence~\cite{kanade2020learning,feng2020codebert} to a short one, such as 512 tokens. But obviously, a 512-long token sequence will lose a significant amount of information in the program or code snippet.

The hierarchical architecture of OSCAR is designed to better solve this problem. We partition the instructions of the input program into groups by every $K$ instructions as one group, and the IR (or abstract environment information) tokens of each group would be fed into parameter-shared IR (or Env.) encoders separately. The output representations coming from one instruction of the token-level encoders, would be averagely pooled, to aggregate the information at the instruction level. Then, those instruction-level hidden representations will be fed to the Inst. encoder for further feature extraction. We set $K=4$ in our experiments.

Similar to ~\citet{dai2020funnel}, we up-sample the output sequences of the instruction-level encoder by repeating each hidden vector multiple times so that the length is enlarged to the original token sequence. After up-sampling, consecutive vectors of each instruction would be exactly the same and lost the detailed token-level signals. To involve the uncompressed token-level signal, we adopt a residual connection between uncompressed token-level hidden representations and the up-sampling vectors. After that, another two token-level encoders would try to recover the original token sequences on the positions of the instruction masks. 

\subsubsection{Positional Condition Encoding}

Since the Transformer is developed to solve the problem of sequence transduction in natural language, it cannot well capture the complicated control structure of programming languages, such as iteration logic and selection logic. However, the control flow information is indispensable for understanding the semantics of a program. To overcome this problem, incorporating the control flow graph (CFG) into Transformer has been widely adopted in prior works~\cite{hellendoorn2019global, david2020neural}. 

\begin{figure}[h]
	\centering
	\includegraphics[width=0.5\textwidth]{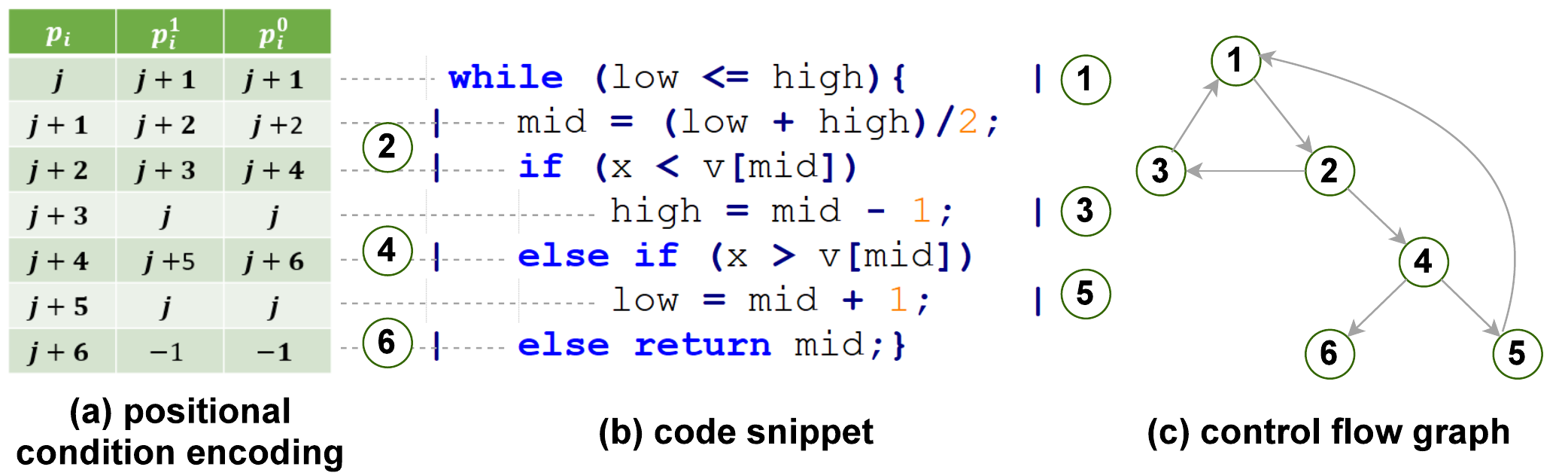}
	\vskip -0.05in
	\caption{\textbf{An illustration of PCE}. PCE could encode the information of control flow graph into the model. Please note that the example code snippet is written in C++ for readability.}\label{fig:pce}
\end{figure}

In this paper, we design a more simple but effective method called Positional Condition Encoding (PCE), to encode the control flow information into the model through positional encoding. PCE assigns three learnable embedding vectors to the position of each instruction in the target program or code snippet, representing the instruction's current position, and target positions after conditionally jumping with true and false, respectively. Fig.\ref{fig:pce} shows the illustration of PCE corresponding to the code snippet and the control flow graph, where $p_i, p^{1}_i$ and $p^{0}_i$ denote the learnable embedding at the current position, true-jumping position, and false-jumping position, of the instruction at position $i$ separately.

To be more concrete, let $h_i\in \mathbb{R}^d$ be the instruction-level hidden representation at position $i$, and $W^V\in \mathbb{R}^{d\times d}$ denotes learnable projection matrix. The output $z_i$ of the first self-attention module in the Inst. encoder can be written as 
\begin{eqnarray}\label{eqn:self-attn}
z_i=\sum_{j=1}^n \frac{\exp (\alpha_{ij})}{\sum_{j'=1}^n\exp (\alpha_{ij'})}(h_jW^{V}).
\end{eqnarray}
Similar to~\citet{ke2020rethinking}, we propose to model the relationship between positions with different projection matrices. Then, the correlation term $\alpha_{ij}$ in Eq.\ref{eqn:self-attn} is calculated as
{\small
\begin{eqnarray}\label{eqn:coff}
    \alpha_{ij}=\frac{1}{\sqrt{4d}} (h_iW^{Q})(h_jW^{K})^T + \frac{1}{\sqrt{4d}} (p_iU^{Q})(p_jU^{K})^T \nonumber \\ 
    + \frac{1}{\sqrt{4d}} (p^{1}_iU^{1,Q})(p^{1}_jU^{1,K})^T + \frac{1}{\sqrt{4d}} (p^{0}_iU^{0,Q})(p^{0}_jU^{0,K})^T,
\end{eqnarray}
}
where $W^Q,W^K\in \mathbb{R}^{d \times d}$ are the projection matrices for the hidden representation $h$, $ U^Q, U^K\in \mathbb{R}^{d \times d}$ are the projection matrices for the current positional embedding $p$, and $U^{1,Q},U^{1,K},U^{0,Q},U^{0,K} \in \mathbb{R}^{d \times d}$ are for the true-jumping and false-jumping position embedding $p^1$ and $p^0$. The scaling coefficient $\frac{1}{\sqrt{4d}}$ maintains the magnitude of $\alpha_{ij}$.

From Fig.\ref{fig:pce} we can see that PCE can incorporate the information about outgoing edges of the nodes in the CFG into the attention module, and the information about incoming edges would also be captured after the calculation of positional correlation in Eq.\ref{eqn:coff}. This indicates that OSCAR could capture all information of the CFG with PCE even that the CFG has not been explicitly fed into the model.

\begin{table*}[htb]
\caption{\textbf{Binary code similarity detection using the \textit{Recall at position 1 (Recall@1)} metric on popular software across different optimization levels.}}
\label{tab:bindiff}
\begin{center}
\begin{tabular}{c|l |cccc|c}
\toprule
Software & Methods & O0-O1 & O0-O3 & O1-O3 & O2-O3 & Avg.\\
\midrule
\multirow{4}{*}{SQLite} & BinDiff~\cite{dullien2005graph} & 0.4360 & 0.0419 & 0.1600 & 0.6455 & 0.3209 \\
\cline{2-7}
& Asm2vec~\cite{ding2019asm2vec} &  0.2407 & 0.2084 & 0.4270 & 0.5520 & 0.2371 \\
\cline{2-7}
& BinaryAI~\cite{yu2020order,yu2020codecmr} & \textbf{0.8245} & 0.5563 & 0.6667 & 0.8107 & 0.7146 \\
\cline{2-7}
& OSCAR & 0.8063 & \textbf{0.6467} & \textbf{0.7148} & \textbf{0.8198} & \textbf{0.7469} \\
\hline
\hline
\multirow{4}{*}{zlib} & BinDiff~\cite{dullien2005graph} & 0.7143 & 0.1237 & 0.1959 & 0.4271 & 0.3653 \\
\cline{2-7}
& Asm2vec~\cite{ding2019asm2vec} & 0.1805 & 0.2371 & 0.3814 & 0.5104 & 0.3274 \\
\cline{2-7}
& BinaryAI~\cite{yu2020order,yu2020codecmr} & \textbf{0.9023} & 0.6392 & 0.7010 & 0.7708 & 0.7533 \\
\cline{2-7}
& OSCAR & \textbf{0.9023} & \textbf{0.7423} & \textbf{0.7835} & \textbf{0.8229} & \textbf{0.8128} \\
\hline
\hline
\multirow{4}{*}{Libcurl} & BinDiff~\cite{dullien2005graph} & 0.5464 & 0.1893 & 0.4831 & 0.8190 & 0.5095 \\
\cline{2-7}
& Asm2vec~\cite{ding2019asm2vec} & 0.4911 & 0.4916 & 0.6012 & 0.6426 & 0.5566 \\
\cline{2-7}
& BinaryAI~\cite{yu2020order,yu2020codecmr} & 0.8550 & 0.7282 & 0.7991 & \textbf{0.8620} & 0.8111 \\
\cline{2-7}
& OSCAR & \textbf{0.8560} & \textbf{0.7405} & \textbf{0.8190} & 0.8512 & \textbf{0.8167} \\
\hline
\hline
\multirow{4}{*}{BusyBox} & BinDiff~\cite{dullien2005graph} & 0.5364 & 0.2939 & 0.6304 & \textbf{0.9658} & 0.6066 \\
\cline{2-7}
& Asm2vec~\cite{ding2019asm2vec} & 0.3236 & 0.3767 & 0.6163 & 0.6907 & 0.5018 \\
\cline{2-7}
& BinaryAI~\cite{yu2020order,yu2020codecmr} & 0.8541 & 0.7907 & \textbf{0.9023} & 0.9478 & 0.8737 \\
\cline{2-7}
& OSCAR & \textbf{0.8764} & \textbf{0.8183} & 0.8883 & 0.9520 & \textbf{0.8838} \\
\hline 
\hline
\multirow{4}{*}{LibTomCrypt} & BinDiff~\cite{dullien2005graph} & 0.1096 & 0.0257 & 0.1768 & 0.6956 & 0.2519 \\
\cline{2-7}
& Asm2vec~\cite{ding2019asm2vec} & 0.4345 & 0.4319 & 0.6869 & 0.7454 & 0.5747 \\
\cline{2-7}
& BinaryAI~\cite{yu2020order,yu2020codecmr} & 0.4906 & 0.4835 & 0.6114 & 0.7491 & 0.5837 \\
\cline{2-7}
& OSCAR & \textbf{0.6483} & \textbf{0.5404} & \textbf{0.6630} & \textbf{0.7583} & \textbf{0.6525} \\
\bottomrule
\end{tabular}
\end{center}
\end{table*}

\subsection{Pre-training Objectives}\label{sec:3.3}

\textbf{Masked Instruction LM} \quad Predicting the masked tokens is the most commonly used objective function in previous BERT-based code representation methods~\cite{kanade2020learning,feng2020codebert,guo2020graphcodebert}. It's essential to OSCAR that captures the token-level contextual knowledge from optimizing MLM loss during pre-training. However, since both IR and abstract environment information are simultaneously provided to our model, it's trivial to derive particular tokens in the IR through the environment which comes from the same instruction, and vice versa. To prevent such potential information leakage, we propose to mask consecutive tokens of an entire instruction. Specially, we sample randomly 15\% instructions from IR and paired environment. We replace the instructions with {\tt[MASK]} tokens 80\% of the time, with random instructions 10\% of the time, and leave them unchanged 10\% of the time.

\textbf{Contrastive Learning with Optimization Techniques} How to effectively capture the program- or code snippet-level semantic knowledge during pre-training is certainly essential for code representation models. However, it has not been well-studied by prior works.

Actually, modern compilers support versatile compilation options for different demands of optimizations, e.g., minimize execution time, memory footprint, storage size, etc. A single source code snippet could be translated to contrasting IR with different optimization techniques, but the meaning of the code would not be changed. Naturally, the different combinations of multiple optimizations can be used as a method of data augmentation for source code (Details in Appendix E). Motivated by this, we propose to employ an objective on {\tt[CLS]} token of contrastive learning with a momentum encoder~\cite{he2020momentum} as OSCAR's self-supervised task to better facilitate the semantics understanding from program level, which is illustrated in Fig.\ref{fig:cros}.

\section{Experiments}

We conduct the pre-training of OSCAR on a large corpus of real-world programs from publicly available open-source GitHub repositories, which covers a broad range of disciplines from operating systems and compilers, to machine learning systems and linear algebra subprograms (Details in Appendix F.1). We evaluate the performance of OSCAR on several semantics understanding tasks for programs in this section. We first perform our model on a practical and important software engineering task, i.e., binary diffing. It is a very fundamental task in reverse engineering and has been widely used to enable different kinds of critical security analysis. After that, we evaluate the performance of OSCAR for high-level PL understanding on the algorithm classification task. Furthermore, as a pre-training method, we investigate the performance of OSCAR in zero-shot learning, where the parameters of OSCAR are fixed. Finally, we analyze the components of our model in the ablation study. Unless otherwise specified, all experiments are conducted on a 12-layer OSCAR model which is composed sequentially of three token-level encoder layers, six instruction-level encoder layers, and three token-level encoder layers. We follow RoBERTa-base~\cite{liu2019roberta} to set other model configurations (Details in Appendix B), e.g., the dimensionality of hidden representation $d$ is set to 768. The total sequence length of Inst. encoder is set to 512, where the IR and Env. encoders each account for 256 instructions. Detailed descriptions of all downstream datasets and optimization strategies of pre-training and fine-tuning could be found in Appendix G and H respectively.

\subsection{Binary Diffing}
Binary code differential analysis, a.k.a. binary diffing, is a fundamental analysis capability, which aims to measure the function-level similarity between two given binaries. We evaluate the performance of OSCAR on binary diffing by following the setting and the dataset described in~\citet{ding2019asm2vec}. In addition to Asm2vec~\cite{ding2019asm2vec}, we further compare OSCAR with two baseline techniques: BinDiff~\cite{dullien2005graph}, which is the de facto standard binary diffing tool based on graph isomorphism detection; and BinaryAI~\cite{yu2020order,yu2020codecmr}, which is a most recently proposed binary code feature extractor based on a hybrid model of neural network with BERT, CNN and GNN, and achieves state-of-the-art performance on code similarity detection.

Following ~\citet{ding2019asm2vec}, we evaluate baseline techniques and OSCAR on five commonly used programs using \textit{Recall@1}. All five programs are compiled with GCC 7.5.0 against four different optimization levels. The results are given in Tab.\ref{tab:bindiff}. As shown, OSCAR consistently outperforms BinDiff, Asm2vec, and BinaryAI across all optimization levels of five programs in terms of recall, by a large margin. For example, in the most difficult matching situation, i.e. diffing between the O0 and O3 optimization levels, OSCAR improves the recall over all baseline techniques on every program.

\subsection{Algorithm Classification}
In this subsection, we study the performance of OSCAR on high-level programming language understanding. We conduct the experiments on POJ-104 dataset~\cite{poj104}, which contains 104 algorithm problems that were submitted to an online judge system.  All samples were written in C/C++ by students. The dataset has around 500 samples per algorithm. The experimental setting we used is exactly same with ProGraML~\cite{cummins2020programl,cummins2020deep}, which achieves state-of-the-art classification accuracy on this dataset.

\begin{table}[h]
\caption{\textbf{Classification error on POJ-104 test dataset.} The performance of all baseline methods is cited from~\citet{cummins2020deep}.}
\label{tab:poj}
\begin{center}
\begin{small}
\renewcommand\tabcolsep{2.8pt}
\begin{tabular}{l|c}
\toprule
Methods & Error(\%) \\
\midrule
TBCNN~\cite{poj104} & 6.00\\
NCC~\cite{ben2018neural} & 5.17\\
XFG~\cite{ben2018neural} & 4.56\\
XFG w/o inst2vec vocab & 4.29\\
ProGraML~\cite{cummins2020programl,cummins2020deep} & 3.38 \\
\midrule
OSCAR & \textbf{1.92}\\

\bottomrule
\end{tabular}
\end{small}
\end{center}
\end{table}

Tab.\ref{tab:poj} shows the results of classification error. According to the table, our model achieves significant improvement comparing with all previous methods by a large margin, which indicates that OSCAR could well understand the semantics of source code written in high-level PLs.

\subsection{Zero-Shot Learning}

In the previous subsection, we show that after fine-tuning the parameters on downstream tasks, OSCAR could outperform prior methods on both binary code or high-level programming language. In this subsection, we further investigate the performance of pre-trained OSCAR in the zero-shot learning setting, i.e., evaluate OSCAR without modifying the parameters. In the comparison, we choose CodeBERT~\cite{feng2020codebert} as a baseline which shows promising zero-shot ability in the PL-NL probing task. We conduct the empirical study on the code similarity task by leveraging the POJ-104 dataset described above. Following~\citet{ye2020misim}, we label two programs as similar if they are solutions to the same problem, and use mean average precision (MAP) as the evaluation metric. The difference is that we only evaluate our model on the \textit{testing} dataset without using the \textit{training} and \textit{validation} sets. 
\begin{table}[htb]
\caption{\textbf{Mean average precision (MAP) on POJ-104 test dataset.} The performance of all baselines is cited from~\citet{ye2020misim}.The pre-trained model of \dag \  is downloaded from the official release of~\citet{feng2020codebert}.}
\label{tab:zeroshot}
\begin{center}
\begin{small}
\renewcommand\tabcolsep{2.8pt}
\begin{tabular}{c|l|c}
\toprule
 & Methods &  MAP(\%) \\
\midrule
\multirow{6}{*}{Trianing-based} & code2vec~\cite{alon2019code2vec} & 1.90 \\
 & NCC~\cite{ben2018neural} & 39.95 \\
 & NCC w/o inst2vec & 54.19 \\
 & Aroma-Dot~\cite{luan2019aroma} & 52.09 \\
 & Aroma-Cos & 55.12 \\
 & MISIM~\cite{ye2020misim} & 82.45 \\
\midrule
\multirow{4}{*}{\tabincell{c}{Pre-training \\ w/o fine-tuning}} & CodeBERT-{\tt[CLS]}~\cite{feng2020codebert}\dag & 10.38 \\
& CodeBERT-avg. of outputs\dag & 9.62 \\
 & OSCAR$_{1-6-1}$  & 45.24 \\
  & OSCAR  & \textbf{49.17} \\
\bottomrule
\end{tabular}
\end{small}
\end{center}
\end{table}

Since there is no supervision on {\tt[CLS]} token in CodeBERT during pre-training, it's potentially unfair to only use the representation on {\tt[CLS]} token in this task. Following \citet{reimers2019sentence}, we additionally calculate the average of the outputs on all tokens of CodeBERT as the representation in comparison. Furthermore, despite that both CodeBERT and OSCAR have 12 transformer layers, OSCAR has more parameters (163M) than CodeBERT (125M) since there are two simultaneous token-level encoders, i.e., IR encoder and Env. encoder. For a fair comparison, we also report the MAP of a shallow OSCAR with only one token-level encoder layer before and after the six instruction-level encoder layers, which is called OSCAR$_{1-6-1}$ and has only 107M parameters.

As shown in Tab.\ref{tab:zeroshot}, without further modifying the parameters, the pre-trained OSCAR and OSCAR$_{1-6-1}$ both show promising performance on code similarity detection, comparing to other pre-trained models. This indicates that OSCAR has the potential of transferability on downstream tasks without fine-tuning. 

Please note, although OSCAR optimizes a similarity loss function (See Sec.\ref{sec:3.3}) in the pre-training phase, the definitions of two data samples as similar are totally different between pre-training and this task: one labels two IRs as similar if they are generated from the same code snippet with different optimization techniques, and the other labels two programs written by different students as similar if they are solutions to the same OJ problem. Therefore, the objectives of OSCAR pre-training and this task are not the same, and the pre-trained OSCAR model demonstrates the capability of semantics understanding of program in the zero-shot learning setting.

\subsection{Ablation Study}
In this subsection, we investigate the effects of each component in OSCAR on binary diffing task using BusyBox, and the experimental setting is identical to above.
\begin{table}[h]
\caption{\textbf{Ablation study on the components of OSCAR.} }
\label{tab:abl}
\begin{center}
\begin{small}
\begin{tabular}{l|c}
\toprule
Methods & Avg. Recall@1 \\
\midrule
OSCAR & \textbf{0.8838}\\
\quad w/o PCE & 0.8662\\
\quad w/o contrastive loss & 0.8267\\
CuBERT~\cite{kanade2020learning} w/ IR & 0.4650 \\

\bottomrule
\end{tabular}
\end{small}
\end{center}
\end{table}
Tab.\ref{tab:abl} ablates the effects of the two components of OSCAR: contrastive loss and PCE. As shown in the figure, all components are beneficial, improving the recall on the binary diffing task. Meanwhile, we further train a BERT on IR corpus, which is similar to CuBERT~\cite{kanade2020learning} because they share exactly the same architecture, and the only difference is that CuBERT is pre-trained on Python corpus. The result shows that, CuBERT with IR performs not well on the binary diffing task, which reflects the hierarchical architecture of OSCAR is also significantly beneficial.

\section{Discussion}
In this section, we discuss a few potential drawbacks of our method, which are left for future work.

\textbf{Real-time Code Analysis} \quad Currently, we analyze the target code snippet relying on compiler and static program analysis, which requires that the target code snippet should be compilable. This dependence may limit the applications of OSCAR on real-time code analysis, such as in the modern integrated development environment (IDE). However, there are many alternatives to choose for real-time IR translation and environment information extraction. For example, the interpreter can translate interpreted language (e.g., Python interpreter) into IR in an interactive style; and even for some compiled languages, interactive interpreters are also been developed (e.g., Cling\footnote{\url{https://github.com/root-project/cling}.} for C++) which support just-in-time (JIT) compilation. With these technologies, there is no need to require the target code snippet to be a compilable program, but only a complete basic block.

\textbf{Token Semantics Analysis of Source Code} \quad When compilers translate source code to IR, partial semantics of tokens is lost since all variables' names would be automatically normalized and replaced by LLVM value identifier. It may lead to a failure of semantics analysis as important information is contained in the variable name. For example, CuBERT~\cite{kanade2020learning} claims that it can detect the following code written in Python is buggy:
{\small
\begin{lstlisting}
num_batches = batch_size / num_examples
\end{lstlisting}
}
where OSCAR may fail in handling this case with high probability. It may be well-solved by keeping the original tokens in IR. We leave it for future work.

\section{Conclusion}

In this paper, we propose a novel pre-training model called OSCAR to learn better code representation. Motivated by operational semantics, we suggest that, instead of learning representation directly from high-level programming languages, the intermediate representation is a better abstraction of the semantics of instructions; meanwhile, to well understand the meaning of a program, we propose the abstract environment information should be necessarily considered. Besides, we introduce two additional techniques to make up the OSCAR. First, we incorporate the control flow information into the model through a novel positional encoding called PCE. Second, to provide a code snippet-level self-supervision during pre-training, we introduce contrastive loss by generating syntactically diverse but functionally equivalent IRs with different optimization techniques. OSCAR empirically shows promising results on practical software engineering tasks, including both binary code and high-level programming language understanding, and also demonstrates the transferability on downstream tasks without modifying the parameters of the pre-trained model.

{\small
\bibliography{ref}
\bibliographystyle{icml2021}
}

\clearpage
\appendix

\section{Details of Abstract Environment Information}

We develop a LLVM pass to produce the environment information. In the environment information, arguments of the functions are named $a_i, i=0,1,2,...,N_{\mathrm{Arguments}}-1$; SSA values are named $v_i, i=0,1,2,...,N_{\mathrm{Values}}-1$; stack variable allocated by \textit{alloca} instruction are named $m_i, i=0,1,2,...,N_{\mathrm{StackVariables}}-1$. There may be one or more constraints of environment for every instruction, and constraints are separated by semicolon. 
\subsection{Constraints of Arithmetic and Logical Operations}
Constraints of arithmetic and logical operations can be represented as follows:
$v_i=x\;\mathrm{op}\; y$ or $v_i=\mathrm{op}\; x$, where $v_i$ is the SSA value of the result of the operation, $x, y$ can be SSA values, arguments or constants, and $\mathrm{op}$ can be binary or unary operators such as $+,-,*,/$ or $\mathrm{trunc}, \mathrm{fptoint}$.
\subsection{Constraints of Memory Operations}
For allocating memory for stack variables, the constraint can be represented as
$v_i = \mathrm{reference}\;m_j$, where $v_i$ is the address of the allocated stack variable $m_j$.

For memory load, the constraint can be represented as 
$v_i \leftarrow m_j=y_0,y_1,...$ or $v_i \leftarrow \mathrm{dereference}\; x=y_0,y_1,...$, where $v_i$ is loaded value and $m_j$ is the loaded stack variable name if the memory address points to a stack variable, otherwise $x$ is the memory address to load. $y_0, y_1,...$ are possible values loaded from the memory address analyzed by \textit{MemorySSA} pass.

For memory store, the constraint can be represented as $v_i \rightarrow m_j$ or $v_i \rightarrow \mathrm{dereference}\; x$, where $v_i$ is the value to store in the memory and $m_j$ is the stack variable to be stored if the memory address points to a stack variable, otherwise $x$ is the memory address, which can be either SSA values, arguments or constants.

For getting element pointer, the constraint can be represented as
$v_i = \mathrm{gep}\;x\;y_0,\;y_1, ...$, where base pointer $x$ can be SSA values, arguments, stack variables or constants, and indices $y_i$ can be $v_i, a_i$ or constants.

\subsection{Constraints of Selection}
For PHI node, the constraint can be represented as
$v_i = x_0,x_1,...$, where $x_0,x_1,...$ are possible values of $v_i$.

For selecting value by condition value, the constraint can be represented as
$v_i = \mathrm{select}\;x\;y_0\;y_1$, where $x$ is a boolean condition value, and $y_0, y_1$ are selected values when $x$ is $\mathrm{true}$ or $\mathrm{false}$ respectively.

\subsection{Other Constraints}
For return value, the constraint can be represented as $\mathrm{ret}=x$, which means that the return value of the function is $x$.

For loop depth, the constraint can be represented as $\mathrm{loop}=0,1,2,...$, which shows the loop depth of the current instruction analyzed by \textit{LoopInfo} pass.

\section{Details of Architecture}

We have two separated positional condition encoding for IR and Env.. For three kinds of IR encoding, there is a special code for {\tt [CLS]}. And for true encoding and false encoding, there is a special code $-1$ for the unknown position. The unknown code is for the instructions like \textit{switch} or \textit{call}, for which we cannot decide the next position or there are more than two target positions. A \textit{switch} instruction can also be converted to a sequence of branches to prevent unknown position code in this case. For Env. encoding, it is similar except that {\tt [CLS]} is replaced by {\tt [SEP]}. The hidden state of {\tt [CLS]} in the last layer of the instruction-level transformer is connected to a MoCo head. The dimension of the MoCo head is 256 and the length of the MoCo queue is 65536. Finally, when applying masked language model, an IR instruction and its corresponding Env. constraints won't be masked at the same time.

\section{The influence of $K$}
In the section of experiments, we set $K=4$ as constant, which means that each IR and Env. Transformer encoder would process sequences with a length of $32$ and $16$ tokens. Larger $K$ would lead to significant computation increment and memory consumption since the complexity of attention layers is quadratic (i.e., $\mathcal{O}(L^2)$). But in the meanwhile, larger $k$ would also improve the capability of capturing the contextual information among long sequences. In this section, we investigate the performance gap between different choices of $K$ in Table.\ref{tab5}. 

\begin{table}[h]
\caption{\textbf{Classification error on POJ-104 test dataset.} ASTNN~\citep{zhang2019novel} could access the symbol names in source code, which will be normalized in other IR-based methods.}\label{tab5}
\begin{center}
\begin{small}
\renewcommand\tabcolsep{2.8pt}
\begin{tabular}{l|c}
\toprule
Methods & Error(\%) \\
\midrule
ASTNN ~\cite{zhang2019novel} & 1.8\\
OSCAR ($K=4$) & 1.92\\
OSCAR ($K=16$) & \textbf{1.72}\\

\bottomrule
\end{tabular}
\end{small}
\end{center}
\end{table}

\section{Hardware-related Program Semantics Understanding}
In this section, we investigate whether OSCAR performs well on hardware-related semantics understanding. We conduct experiments on two widely-used tasks: device mapping and coarsening threads predictions. We exactly follow the same experimental settings with ~\citep{ben2018neural,cummins2020deep}. The results have shown in Table.\ref{tab6} and \ref{tab7}. In both experiments, OSCAR performs well comparing to baseline methods, and shows good capabilities of program semantics understanding on hardware-related tasks.

\thispagestyle{empty}
{
\begin{table}[th]
\footnotesize
    \centering
    \caption{Error rate (\%) on device mapping task.}\label{tab6}
    \begin{tabular}{c|c|c}
        & AMD & NVIDIA \\
        \hline
        DeepTune$_{IR}$ & 26.9 & 31.6 \\
        inst2vec\cite{ben2018neural} & 19.7& 21.5\\
        ProGraML\cite{cummins2020deep} & 13.4& 20.0\\
        OSCAR & \textbf{11.2}& \textbf{10.3}\\
        \hline
    \end{tabular}

    \label{tab:1}
\end{table}
\begin{table}[th]
\footnotesize
    \caption{Speedups achieved by coarsening threads}\label{tab7}
    \centering
    \begin{tabular}{c|c|c|c|c}
         & DeepTune$_{IR}$ & inst2vec & inst2vec-imm & OSCAR \\
        \hline
         Cypress  & 1.17 & \textbf{1.37} & 1.28 & 1.35\\
         Tahiti & 1.23 & 1.10 & 1.18 & \textbf{1.30}\\
         Fermi & 1.14 & 1.07 & 1.11 & \textbf{1.27}\\
         Kapler & 0.93 & 1.06 & 1.00 & \textbf{1.12}\\
         \hline
    \end{tabular}
    \label{tab:2}

\end{table}
}

\section{Compliant Options for Constrative Learning}
We totally generate 19 variants for every function from different sequences of LLVM passes. Firstly, we generate three variants using \textit{opt} of the LLVM toolchain with standard passes of -O1/2/3. Then for every LLVM IR assembly file, we randomly drop and shuffle the passes of the -O2 optimization level and use \textit{opt} to generate the variants. The standard -O2 optimization passes are shown in Tab.\ref{tab:o2passes}. The algorithm for generating the passes is as follows:

\begin{algorithm}[H]
\label{algo:passes}
\SetAlgoLined
\SetKw{KwBy}{by}
\KwIn{List of the standard -O2 optimization passes $P$, maximal shuffled items $M$ which is even.}
\KwOut{List of the generated optimization passes $P'$}
 Generate a random integer $N\in[0,\mathrm{len}(P)]$;
 
 Randomly select $N$ items $P'$ from $P$;
 
 Generate a random even integer $m\in[0, M]$;
 
 Randomly select $m$ unique items $S$ from $\{0,1,...,N-1\}$;
 
 \For{$i\gets 0$ \KwTo $m-2$ \KwBy $2$}{
  $P'[S[i]] \leftrightarrow P'[S[i+1]]$;
 }
 \caption{Generating LLVM passes}
\end{algorithm}

In our case, we set M = 20.

\begin{table}[h]
\caption{-O2 optimization passes}
\label{tab:o2passes}
\begin{center}
\begin{small}
\renewcommand\tabcolsep{2.8pt}
\begin{tabular}{ll}
\midrule
-tbaa & -scoped-noalias \\
-forceattrs & -inferattrs \\
-ipsccp & -called-value-propagation \\
-attributor & -globalopt \\
-mem2reg & -deadargelim \\
-instcombine & -simplifycfg \\
-prune-eh & -functionattrs \\
-sroa & -early-cse-memssa \\
-speculative-execution & -jump-threading \\
-correlated-propagation & -simplifycfg \\
-domtree & -instcombine \\
-libcalls-shrinkwrap & -pgo-memop-opt \\
-tailcallelim & -simplifycfg \\
-reassociate & -loop-simplify \\
-lcssa & -scalar-evolution \\
-loop-rotate & -licm \\
-loop-unswitch & -simplifycfg \\
-instcombine & -loop-simplify \\
-lcssa & -scalar-evolution \\
-indvars & -loop-idiom \\
-loop-deletion & -loop-unroll \\
-mldst-motion & -phi-values \\
-gvn & -phi-values \\
-memcpyopt & -sccp \\
-demanded-bits & -bdce \\
-instcombine & -jump-threading \\
-correlated-propagation & -phi-values \\
-dse & -loop-simplify \\
-lcssa & -scalar-evolution \\
-licm & -adce \\
-simplifycfg & -instcombine \\
-barrier & -elim-avail-extern \\
-rpo-functionattrs & -globalopt \\
-globaldce & -float2int \\
-lower-constant-intrinsics & -loop-simplify \\
-lcssa & -scalar-evolution \\
-loop-rotate & -loop-distribute \\
-scalar-evolution & -demanded-bits \\
-loop-vectorize & -loop-simplify \\
-scalar-evolution & -loop-load-elim \\
-instcombine & -simplifycfg \\
-scalar-evolution & -demanded-bits \\
-slp-vectorizer & -instcombine \\
-loop-simplify & -lcssa \\
-scalar-evolution & -loop-unroll \\
-instcombine & -loop-simplify \\
-lcssa & -scalar-evolution \\
-licm & -transform-warning \\
-alignment-from-assumptions & -strip-dead-prototypes \\
-globaldce & -constmerge \\
-loop-simplify & -lcssa \\
-scalar-evolution & -loop-sink \\
-instsimplify & -div-rem-pairs \\
-simplifycfg \\
\midrule
\end{tabular}
\end{small}
\end{center}
\end{table}

\section{Pre-training Data and Pre-Processing}

\subsection{Pre-training Data}\label{sec:3.4}

We assembled a large corpus of real-world programs for pre-training from publicly available open-source non-fork GitHub repositories, summarized in Table.\ref{tab:pretraindata}. The software covers a broad range of disciplines from operating systems and compilers, to machine learning systems and linear algebra subprograms. After collecting the corpus, we first compile them into LLVM IRs using Clang 11 with -O0 optimization level\footnote{Except for Linux Kernels which could only be built with -O1 or above.}. Then, for each program, we further generate 19 variants with the same functionality (20 in total), by random arrangement and combination of different LLVM optimization passes. After that, we sample about 500k functions from the dataset. In the pre-training phase, we sample several functions from the dataset to form a mini-batch as the training data for each iteration.

\begin{table}[h]
\caption{The eleven sources of LLVM IR used to produce pre-training dataset. All software is downloaded from Github.}
\label{tab:pretraindata}
\begin{center}
\begin{small}
\renewcommand\tabcolsep{2.8pt}
\begin{tabular}{lc|rr}
\toprule
Software & Domain & \#instructions & \#functions\\
\midrule
Linux-vmlinux & Linux Kernel & 2,930,372 & 45,368 \\
Linux-modules & Linux Kernel & 16,509,892 & 229,942 \\
GCC & Compiler & 1,816,782 & 22,383 \\
MPlayer & Multimedia & 1,223,068 & 12,747 \\
OpenBLAS & BLAS & 515,985 & 5,415 \\
PostgreSQL & Database & 939,199 & 12,807 \\
Apache & Web Server & 390,135 & 5,519 \\
Blender & 3-D Creation & 5,925,801 & 123,689 \\
ImageMagcick & Image Processing & 440,265 & 7,182 \\
Tensorflow & Machine Learning & 12,041,852 & 294,553 \\
Firefox & Browser & 5,290,430 & 96,187 \\
\midrule
Total & & 48,023,781 & 855,792\\
\bottomrule
\end{tabular}
\end{small}
\end{center}
\end{table}

\subsection{Pre-Processing}

Firstly, we use wllvm\footnote{\url{https://github.com/travitch/whole-program-llvm}} with Clang 11 to compile the source code to LLVM IR. For every object file, wllvm will generate an LLVM IR bitcode file, which can be then converted to an LLVM IR assembly file. For every LLVM IR assembly file, we extract the functions which occur in all 20 variants of the file.

Then we use the above-mentioned LLVM pass to filter out the functions which exceed the maximal instructions as well as generate PCE, environment information, and IR instructions with normalized identifier names.

After that, we tokenize the LLVM IR assembly code and process the names of functions and types as follows:
\begin{enumerate}
\item If an identifier name is a mangled C++ symbol, demangle it and remove extra information. Only function names will be retained. Also, for type names, extra information such as template arguments or namespaces will be removed.
\item Split the names into words by underscore and case.
\item Use byte pair encoding to break down the words into subwords.
\end{enumerate}

Literal constants will also be split into subwords using BPE.

Finally, we convert IR instructions and environment information into raw text and split them into the training set and the validation set with the ratio of 19:1.

\section{Downstream Dataset}
\subsection{Binary Diffing}

We collected several programs and libraries. The numbers of the programs in the training/validation/testing dataset are 13, 2, and 5. Firstly, we compile them using GCC 7.5.0 with debug information and four different optimizations levels (-O0/1/2/3). Then, we use debug information to generate the ground truth of matched functions in different variants of binaries and then stripped the debug information out of the binaries as well as replace the function symbols with meaningless strings. We only treat two binary functions as equivalent if their function names and source code filenames in the debug information are both the same. In this way, we can ensure that the ground truth collected is correct, though it may not be exhaustive. After that, we use retdec\footnote{\url{https://retdec.com/}} decompiler to convert the binaries to LLVM IR, and then process the IR to generate raw text input in the above-mentioned way. 

For the training and validation set, only the functions that occur in all four variants of a binary will be used. However, for the test set, all the functions will be included as we need to retrieve a function from all the functions of another binary. The numbers of the functions in the training/validation/testing dataset are 71000, 5804, and 40791.

Before matching the functions using BinDiff\cite{dullien2005graph}, we remove the names of the functions in IDA except for the exported symbols as BinDiff will match two functions if they have the same name, which results in invalid results.

We use \textit{Recall@1} as the evaluation metrics, which can be computed in this manner:

For binaries $B_1$, $B_2$ as the sets of binary functions, we have a ground truth mapping $f_1:B'_1 \rightarrow B'_2$, where $B'_1 \subseteq B_1, B'_2 \subseteq B_2$. For every $x_1 \in B'_1$, we also find a $x_2=f_2(x_1) \in B_2$ which maximizes $\mathrm{similarity}(x_1,x_2)$ computed by our model, which is the cosine similarity of the {\tt [CLS]} feature vectors of these two functions. MoCo\cite{he2020momentum} head is not involved in the computation of the feature vectors. Then, we have:
$$\mathrm{Recall}@1 = \frac{|f_1 \cap f_2|}{|f_1|}$$
\subsection{POJ-104}
POJ-104 dataset\cite{poj104} is collected from an online judge platform, which contains 104 program classes written by 500 different people randomly selected per class, so there are a total of 52000 samples in the dataset. We use the dataset for the task of clone detection and algorithm classification.

For the POJ-104 clone detection task, we compile the code of the POJ-104 dataset to LLVM IR assembly files with Clang 11 and -O0 optimization level. To compile the code successfully, we prepend following statements before the code:
\begin{verbatim}
#include <bits/stdc++.h>
using namespace std;
\end{verbatim}

Then, we replace \textit{void main} with \textit{int main} and disable all the warnings to compile the source code. After that, we extract the IR instructions, environment information , and PCE information from the produced LLVM IR assembly files in the above-mentioned way. We concatenate the functions in an LLVM IR assembly file into a single input sequence and truncate it to 255 instructions.

Finally, we split the dataset according to the labels. 64 classes of programs are used for training; 16 classes of programs are used for validation; 24 classes of programs are used for testing.

For the algorithm classification task, we use the compiled IR files from the dataset processed by NCC\cite{ben2018neural}\footnote{\url{https://github.com/spcl/ncc/tree/master/task}}. The dataset is split by 3:1:1 for training, validation, and testing. To successfully compile the programs, \textit{\#include} statements are also prepended before the source code. Data augmentation is applied on the training set by compiling each file 8 times with different optimization options (-OO/1/2/3 and w/ or w/o -ffast-math). We keep up to four functions per source code file and truncate each function to 255 instructions.

We use MAP@R as the evaluation metrics of the clone detection task. MAP@R is defined as the mean of average precision scores, each of which is evaluated for retrieving R most similar samples given a query. In our case some source code files (\textasciitilde3\%) are not compilable, so we only retrieve $R_i$ most similar samples for every query where $R_i$ is the number of the valid samples of the same class with the query $s_i$. Detailed information of how to compute our evaluation metrics is as follows.

We denote the set of all the samples as $S=\{s_i\;|\;i=0,1,2,...,N-1\}$, where $N$ is the number of the samples. And the label of $s_i$ is $l(s_i)$. Then, we denote the similarity scores between $s_i$ and $s_j$ computed by our model $f$ as $\mathrm{similarity}(s_i, s_j)=\cos (f(s_i),f(s_j))$. The feature vectors $f(s_i)$ and $f(s_j)$ computed by our model are the output of the two-layer MLP of the MoCo head.

For every $s_i\in S$, let $S_i=\{s_j \in S\;|\;l(s_j)=l(s_i), s_j\neq s_i\}$, and $R_i=|S_i|$. We retrieve $R_i$ most similar samples as $Q_i$ from $S-\{s_i\}$ by similarity scores $\mathrm{similarity}(s_i, s_j), s_j\in S-\{s_i\}$. Then, we have:
$$\mathrm{Precision}_i=\frac{|Q_i\cap S_i|}{|S_i|}$$
$$\mathrm{MAP@R}=\frac{1}{N}\sum_{i=0}^{N-1}\mathrm{Precision}_i$$

\section{Training Details}

\subsection{Pre-training}
The loss for the pre-training task is:

$$L=\lambda L_\mathrm{MLM} + \mu L_\mathrm{MoCo}$$ where $\lambda$ is MLM loss coefficient and $\mu$ is MoCo loss coefficient. We strictly follow the algorithm of MoCo, except that $x_{\mathrm{key}}$ is an augmented image in MoCo, while $x_{\mathrm{key}}=[x_{\mathrm{IRkey}}:x_{\mathrm{EnvKey}}]$ is the augmented IR instruction and its environment information  in our model.

We pretrain the model on 8 V100 GPUs with the hyper-parameters shown in Tab.\ref{tab:pretrainhyperparameters}.

\begin{table}[h]
\caption{Hyper-parameters for pre-training.}
\label{tab:pretrainhyperparameters}
\begin{center}
\begin{small}
\renewcommand\tabcolsep{2.8pt}
\begin{tabular}{l|r}
\toprule
Hyper-parameter & Value \\
\midrule
Training steps & 1000000 \\
Warm-up steps & 30000 \\
Peak LR & 0.0001 \\
Batch size & 16 \\
Update frequency & 4 \\
Dropout & 0.1 \\
Attention dropout & 0.1 \\
Weight decay & 0.01 \\
MoCo dimension & 256 \\
MoCo temperature & 0.02 \\
MoCo momentum & 0.999 \\
MoCo queue length & 65536 \\
MLM loss coefficient & 1 \\
MoCo loss coefficient & 1000 \\
\bottomrule
\end{tabular}
\end{small}
\end{center}
\end{table}

\subsection{Binary Diffing}
When training OSCAR for the binary diffing task, we firstly sample a mini-batch of triplets of two samples $v_i, v^+_i$($i=0,1,2,...,N-1$,$N$ is the size of the mini-batch) of the same label, i.e. from the binary functions generated by different optimizations with the same source code function, and one sample of another label $v^-_i (i=0,1,...,N-1)$. The feature vectors of the triplets are denoted $v_0, v^+_0, v^-_0; v_1, v^+_1, v^-_1; ...; v_{N-1},v^+_{N-1},v^-_{N-1}$. The label of $v_i$ is $l(v_i)$, and we have $l(v_i)=l(v^+_i)\neq l(v^-_i)$. The loss $L$ of the mini-batch is computed as follows:
$$\tau=\sqrt{d}=\sqrt{768}$$
$$
p_i=\exp(v_i\cdot v^+_i/\tau)
$$
$$
s_{ij}=\exp(v_i\cdot v_j/\tau)+\exp(v_i\cdot v^+_j/\tau)
$$
$$
n_i=\exp(v_i\cdot v^-_i/\tau)+\sum_{j=0;\;l(v_j)\neq l(v_i)}^{N-1} s_{ij}
$$
$$
L=-\frac{1}{N}\sum_{i=0}^{N-1}\log\frac{p_i}{p_i+n_i}
$$

The feature vectors are the last-layer hidden states of the {\tt [CLS]} tokens in the instruction-level transformer. MoCo head including the two-layer MLP is dropped. We train the model on 4 V100 GPUs for 128000 steps with 6400 warm-up steps. Peak learning rate is 0.00002; weight decay is 0.2; dropout and attention dropout is 0.1; batch size is 48 and update frequency is 1.

We use BinDiff, Asm2vec\cite{ding2019asm2vec}\footnote{\url{https://github.com/McGill-DMaS/Kam1n0-Community}} and BinaryAI\cite{yu2020order,yu2020codecmr}\footnote{\url{https://github.com/binaryai/sdk}} v2 API as the baseline. All hyper-parameters of Asm2vec are default. BinaryAI uses IDA Pro and its Hex-Rays decompiler to generate C-like pseudo-code for binary functions, and then upload it to Tencent's server to compute the similarity of the functions. Also, Asm2vec and BinDiff both depend on IDA Pro and its dissembler or decompiler.
As the Hex-Rays decompiler is considered better than the retdec decompiler, we think that the comparison between OSCAR and BinaryAI is reasonable.

\subsection{Code Classification}

We firstly sum the {\tt [CLS]} vectors of each function in an LLVM IR assembly file to get the representation of the sample. Then the feature vectors are feed into a fully connected layer followed by a projection layer and a softmax layer. After that, we use the cross-entropy loss for the classification task.

We train the model on 8 V100 GPUs for 100000 steps with 10000 warm-up steps. Peak learning rate is 0.00005; weight decay is 0.01; dropout and attention dropout is 0.1; batch size is 8 and update frequency is 4.

\end{document}